# Article Title:

# Deep learning models for price forecasting of financial time series: A review of recent advancements: 2020-2022

## Authors:

| |
|---|
| **Cheng Zhang***  <br> Advanced Informatics Department, Razak Faculty of Technology and Informatics, Universiti Teknologi Malaysia, 54100, Kuala Lumpur, Malaysia  <br> Email: zcheng582dx@gmail.com  <br> ORCID iD: 0000-0002-4150-3371 |
| **Nilam Nur Amir Sjarif**  <br> Advanced Informatics Department, Razak Faculty of Technology and Informatics, Universiti Teknologi Malaysia, 54100, Kuala Lumpur, Malaysia  <br> Email: nilamnur@utm.my  <br> ORCID iD: 0000-0003-4969-9708 |
| **Roslina Ibrahim**  <br> Advanced Informatics Department, Razak Faculty of Technology and Informatics, Universiti Teknologi Malaysia, 54100, Kuala Lumpur, Malaysia  <br> Email: iroslina.kl@utm.my  <br> ORCID iD: 0000-0002-1343-5842 |

**Abstract**

Accurately predicting the prices of financial time series is essential and challenging for the financial sector. Owing to recent advancements in deep learning techniques, deep learning models are gradually replacing traditional statistical and machine learning models as the first choice for price forecasting tasks. This shift in model selection has led to a notable rise in research related to applying deep learning models to price forecasting, resulting in a rapid accumulation of new knowledge. Therefore, we conducted a literature review of relevant studies over the past three years with a view to aiding researchers and practitioners in the field. This review delves deeply into deep learning-based forecasting models, presenting information on model architectures, practical applications, and their respective advantages and disadvantages. In particular, detailed information is provided on advanced models for price forecasting, such as Transformers, generative adversarial networks (GANs), graph neural networks (GNNs), and deep quantum neural networks (DQNNs). The present contribution also includes potential directions for future research, such as examining the effectiveness of deep learning models with complex structures for price forecasting, extending from point prediction to interval prediction using deep learning models, scrutinising the reliability and validity of decomposition ensembles, and exploring the influence of data volume on model performance.

## 1. INTRODUCTION

Financial market forecasting has long been a complex yet crucial task. The potential for financial gains through successful forecasting has driven numerous practitioners to participate in this activity. Among the various types of forecasting tasks in the financial sector, a fundamental and pivotal task is price forecasting or predicting the future values of a financial time series (Sezer et al., 2020). As financial markets continuously evolve and adapt, the need for robust and reliable price forecasts has become increasingly evident. If accurate, price forecasting results are valuable for investors, traders, and financial institutions in making informed decisions, managing risks, and optimising investment strategies (Cavalcante et al., 2016; Tang et al., 2022). However, accurately predicting the prices of financial time series is extremely challenging owing to the intricate and uncertain nature of the data. The efficient market hypothesis (EMH) proposed by Fama (1970) posits that asset prices reflect all available information, implying that the speed at which profitable trading strategies are developed often does not keep pace with price fluctuations, leading to price adjustments before the strategy is implemented. Simultaneously, price behaviour in an efficient market is similar to a random walk, leaving indiscernible patterns in the historical data (Fama, 1995).

Several methods can forecast the prices of financial time series, but each has its own limitations. Fundamental analysis, which relies mainly on a forecaster's domain knowledge to discern useful information from available data, has become less popular because of its low efficiency (Deboeck, 1994; Rouf et al., 2021). Statistical methods, such as autoregressive integrated moving average (ARIMA) and generalised autoregressive conditional heteroskedasticity (GARCH), rely on a predefined mathematical structure and statistical assumptions to model and analyse time series data. However, they may not capture the nonlinear patterns or complex relationships present in real-world financial time series (Adhikari & Agrawal, 2014; Cheng et al.,

2015). Machine learning methods such as artificial neural networks (ANNs) and support vector regression (SVR) can automatically discern patterns hidden in historical data. However, they are dependent on feature engineering and are bottlenecked by model complexity when numerous training data are available (Jiang, 2021). Consequently, there remains a need for more advanced price-forecasting methods.

Deep learning techniques exert a transformative influence within this field. Long short-term memory (LSTM) networks, a well-known variant of recurrent neural networks (RNNs), have shown remarkable success in capturing temporal dependencies between data points in financial time series and have replaced traditional statistical and machine learning methods as the preferred choice for price forecasting tasks (Durairaj & Mohan, 2019; Nosratabadi et al., 2020; Sezer et al., 2020; Hu et al., 2021; Kumar et al., 2021; Lara-Benítez et al., 2021). Convolutional-recurrent neural networks (CRNNs), a type of hybrid deep learning model that leverages the complementary strengths of convolutional neural networks (CNNs) and RNNs to capture spatiotemporal patterns in time series data, have also been widely applied to price forecasting (Zhang et al., 2019; Tsantekidis et al., 2020). One advantage of deep learning-based forecasting models is that they can automatically learn and adapt to intricate patterns while capturing both linear and nonlinear dependencies in the data, relying less on domain knowledge and benefiting from rich training data (Bengio, 2012; Goodfellow et al., 2016). This shift in model selection is also driven by the parallel processing power offered by graphics processing units (GPUs) and the facilitation of implementing and testing deep learning models using high-level programming packages, such as TensorFlow and PyTorch (Jiang, 2021).

Consequently, there has been a notable rise in research related to applying deep learning models to price forecasting. This research area rapidly iterates, indicating that a large quantity of newly generated knowledge is accumulated within a couple of years but lacks adequate exploration. Although several studies have reviewed the application of machine learning and deep learning methods to predict financial asset prices (Gandhmal & Kumar, 2019; Bustos & Pomares-Quimbaya, 2020; Nosratabadi et al., 2020; Nti et al., 2020; Sezer et al., 2020; Jiang, 2021; Thakkar & Chaudhari, 2021), these reviews have predominantly focused on studies published prior to 2020, implying that the latest progress in this domain, particularly the applications of cutting-edge deep learning models for price forecasting in the last three years, may not be included or extensively discussed in these studies. We acknowledge the challenge for researchers and practitioners interested in this field to stay up-to-date and to have information on future research directions. As a response, we conducted a literature review of recent advancements in the application of deep learning models to the price forecasting of financial time series, with a focus on developments from 2020 to 2022.

Overall, there is both homogeneity and specificity in the studies covered in this review. The homogeneity stems from the similar characteristics of the target variables and similar prediction workflows in each forecasting task. In contrast, the specificity stems from the diversity of the proposed deep learning models, each with a unique architecture that reflects a specific modelling approach. Therefore, we place less stress on data collection and preprocessing but more on the details of deep learning-based forecasting models. By offering information on model architectures, applications, advantages, and disadvantages, we aim to assist readers in gaining insight into this rapidly developing interdisciplinary area. Moreover, the future research directions proposed in this review may contribute to long-term advancements in this field.

The remainder of this paper is organised as follows. Section 2 provides the background information on price forecasting. Section 3 provides an overview of the relevant studies that were reviewed. Section 4 introduces the deep-learning forecasting models proposed in the literature. Section 5 outlines potential directions for future research. Finally, Section 6 presents the conclusions of this study.

## 2. FORECASTING BACKGROUND INFORMATION

Before delving into the details of deep learning models for price forecasting of financial time series, we first provide background information on the price-forecasting task. The following subsections explain the concept and characteristics of financial time series, the definition of price forecasting, and the workflow of prediction.

### 2.1 Concept of financial time series

A time series is a finite sequence of data recorded and chronologically indexed (Al-Hmouz et al., 2015). Time series data in various domains are often analysed to uncover patterns, trends, and relationships that evolve over time, enabling people to make informed decisions and predictions for certain tasks. In the financial market, observations of one asset price taken at equally spaced time points over a given period constitute a financial time series (Tsay, 2005). This definition suggests that a price series can refer to "open" or "close" prices depending on whether the sampling time point corresponds to the initial or final time point in a transaction time unit, respectively. Additionally, observations could also be the highest and lowest prices reached by the asset during each transaction time unit, known as the "high" and "low" prices, respectively, or the number of transactions within each transaction time unit, known as the "volume". The close prices are adjusted based on the split multiplier specified in the dividend adjustment to derive "adjusted close" prices. These different types of observations constitute historical trading data. According to the length of the transaction time unit, historical data can refer to tick-level, minute-level, hourly, daily, weekly, monthly, or yearly data.

Financial assets usually refer to stock indices, individual stock prices, commodity prices, cryptocurrency rates, and forex exchange rates (Taylor, 2008). Different types of assets and financial instruments are denominated or quoted in specific ways based on their nature and the markets in which they operate. Stock and commodity prices are denoted by a specific currency, and cryptocurrency rates are usually denoted by other cryptocurrencies or fiat currencies. Forex exchange rates represent the relative value between two currencies, whereas stock indices and funds represent value aggregates within their respective contexts.

Conventionally, a univariate financial time series in a forecasting task consists of observations of a target variable. The most common univariate financial time series analysed by the financial sector is the daily close price series. In contrast, multivariate financial time series include observations of several variables over the same period, and these variables usually have complex interdependencies. The records of technical indicators (e.g., moving averages and relative strength index) or sentiment indicators (e.g., social media sentiment scores) usually do not refer to the observations of a target variable but constitute an exogenous time series that offers supplemental data for enriching the feature set of the model input.

## 2.2 Characteristics of financial time series

Although financial time series are diverse in terms of the underlying assets they track, the unit of measurement, and the value range of each series, they exhibit several common properties, such as trends, seasonality, and random factors. Trends refer to long-term directional movements in a series, such as upward or downward trends, indicating the overall market movement. Seasonality refers to repetitive patterns that occur at regular intervals and are often influenced by calendar effects. Random factors represent unpredictable, nonsystematic components of the series, including noise, shocks, and irregular fluctuations. In addition, financial time series are characterised by volatility, nonlinearity, heteroscedasticity, and autocorrelation, which pose significant challenges for precise modelling and accurate forecasting (Tsay, 2005).

In general, a financial time series that spans a considerable timeframe exhibits nonstationary characteristics such as varying means and variances over time. Instances of strictly stationary financial time series within real-world scenarios are infrequent, occurring only when asset prices maintain a consistent mean and variance over a mature period and their autocorrelation demonstrates relative stability. Traditional statistical methods for time series analysis assume a stable connection between past and future values, a presumption that can falter in the presence of nonstationarity within a financial time series (Mills & Markellos, 2008). Consequently, techniques such as detrending and differencing are frequently employed to reveal stationary tendencies in the data. However, it is noteworthy that stationarity prerequisites need not be obligatory when deploying a deep learning-based forecasting model because the model possesses the capacity to independently discern irregular patterns embedded within the data.

## 2.3 Definition of price forecasting

Price forecasting of financial time series complies with the rule of point forecasting of general time series. Given a time series $y$, whose time steps are given as $k \in [1, K]$, a time series forecasting model $f(\cdot)$ estimates the future value $\hat{y}$ at a forecast horizon $H \geq 1$, utilising the historical values (from time steps $k$ to $k - H_1$, with $H_1$ being the time lags) of the desired time series and several exogenous time series, which are denoted as $y[k], \cdots, y[k - H_1]$ and $\mathbf{u}^T[k], \cdots, \mathbf{u}^T[k - H_1]$, respectively (González Ordiano et al., 2018; Meisenbacher et al., 2022). This functional relationship is defined as follows:

$$\hat{y}[k+H] = f(y[k], \cdots, y[k-H_1], \mathbf{u}^T[k], \cdots, \mathbf{u}^T[k-H_1]; \boldsymbol{\theta}); k > H_1, \qquad (1)$$

where the vector $\boldsymbol{\theta}$ denotes the model parameters.

In the context of price forecasting, the forecasting model $f(\cdot)$ maps an input sequence to a future value of the asset price. The input sequence comprises two subsequences. The first is $y[k], \cdots, y[k - H_1]$, which refers to a univariate sequence extracted from the financial time series that contains observations of the target variable. The second is $\mathbf{u}^T[k], \cdots, \mathbf{u}^T[k - H_1]$, which refers to a multivariate sequence extracted from exogenous time series that contain observations of other related variables, such as technical indicators, sentiment indicators, or macroeconomic variables. Generally, the construction of input sequences for price forecasting is highly empirical.

## 2.4 Workflow of price forecasting

The workflow of price forecasting involves several stages, as illustrated in Figure 1. First, relevant historical raw data are collected, and their quality and consistency are checked. Second, the model inputs are prepared through a series of data processing steps. These steps may involve data denoising, which aims to reduce noise from the raw data, or feature engineering, which aims to extract informative feature sets from data. After these steps, the entire time series, univariate or multivariate, is transformed into short sequences using a sliding window approach in a supervised learning scheme. Fixed-length segments of data are extracted as the window (time lag) moved over the entire series. These segments are then labelled with the corresponding observations of the target variable to train the forecasting models. Third, a forecasting model is selected or developed based on specific requirements. This step is largely determined by the data availability, computational resources, and domain knowledge. Figure 2 illustrates the effect of changing the model complexity and data availability on model performance. Generally, a complex model requires a larger training dataset than a simpler model to achieve optimal performance. Utilising a sizable dataset to train a simple model may lead to high bias or underfitting, whereas using a smaller dataset to train a complex model could result in high variance or overfitting. It is crucial to mention that deep learning models are data-driven, and a complex model does not consistently surpass a simpler model in all forecasting undertakings. Individual domain expertise and data handling inclinations also play roles in the model selection. When the model is trained and its performance is deemed satisfactory after the model evaluation, it can be deployed to make future price predictions.

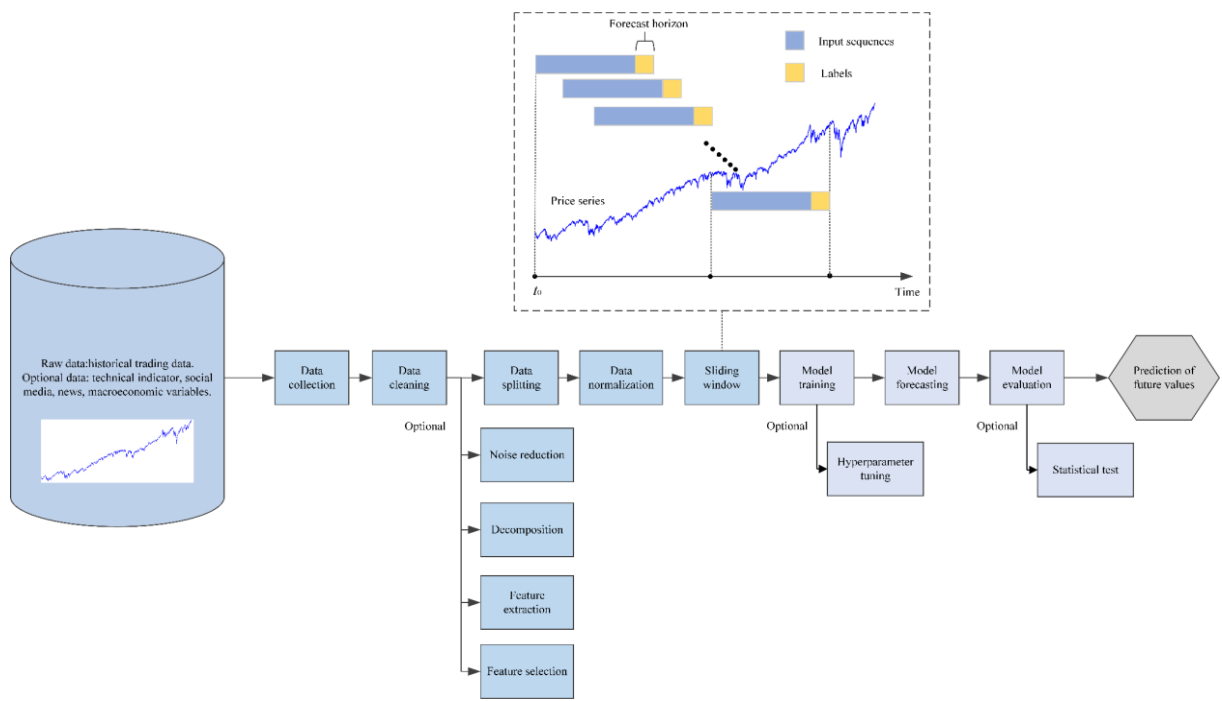

**Figure 1.** Workflow of price forecasting of financial time series.

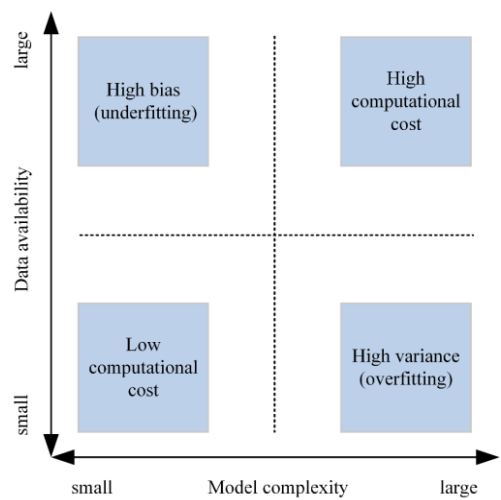

**Figure 2.** Effect of changing model complexity and data availability on model performance.

## 3. OVERVIEW

In this section, we provide an overview of the studies reviewed. The objective of the literature search was to identify studies that utilised deep learning models to predict the prices of financial time series. Therefore, the selection of articles was based on the following criteria: first, the included studies aimed to predict the future price of a financial time series; second, the proposed forecasting models had to be based on deep learning algorithms. To conduct the search, a set of keywords was employed: ("financial time series" and ("deep learning" or "neural network") and (forecast* or predict*) and price). Scopus and Web of Science were chosen as the two scientific databases for sourcing articles published between 2020 and 2022.

The selection process for the included studies is shown in Figure 3. Of 711 articles that met the search criteria, 167 were duplicates. An additional 254 articles were excluded based on the inclusion criteria after screening titles and abstracts. Subsequently, 59 articles were excluded because their full texts could not be retrieved. After a thorough reading of the remaining articles, 157 studies were excluded for the following reasons: 1) unrelated to the price forecasting of financial time series (n=23); 2) unrelated to deep learning (n=65); 3) lack of samples, results, or method details (n=60); 4) comparative studies (n=8); and 5) containing similar content (n=2). As a result of this selection process, 72 articles were included in this review, of which 62 were published in indexed journals.

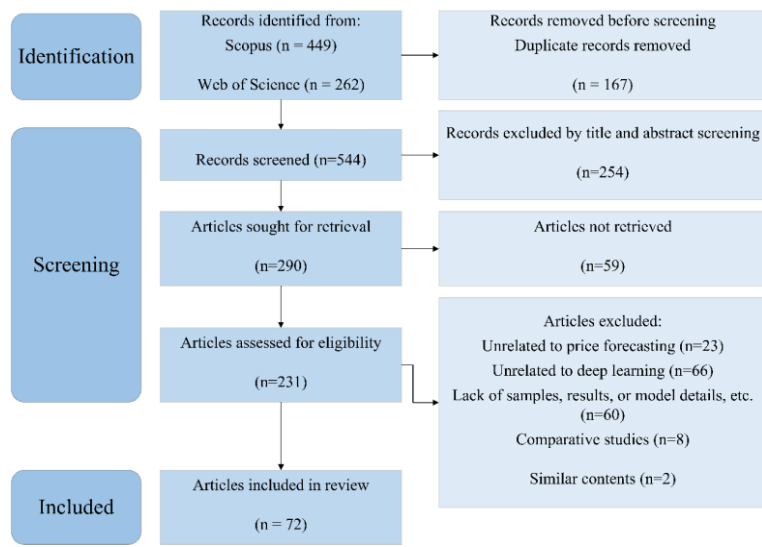

**Figure 3.** Selection process for included studies.

Figure 4 summarises the different types of financial asset prices predicted in the reviewed studies. Notably, the popularity of stock index forecasting has decreased over the past three years, and there has been increasing research interest in forecasting cryptocurrencies and forex rates. Despite these trends, stock market predictions continue to attract more attention from researchers than other types of predictions, and fund forecasting has not yet received significant attention.

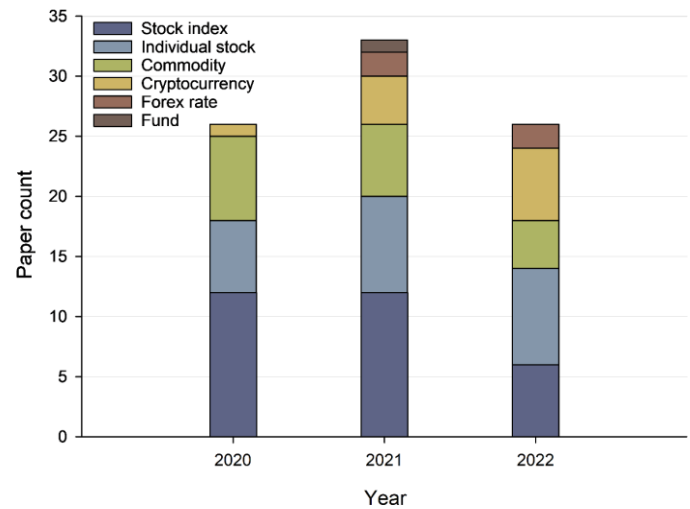

**Figure 4.** Summary of different types of asset prices predicted in reviewed studies.

Figure 5 further summarises the specific financial assets predicted in the reviewed studies. For stock index forecasting, a large proportion of studies have focused on the Standard & Poor's 500 Index (S&P 500) in the United States and the Shanghai Stock Exchange (SSE) Composite in China. Additionally, crude oil and gold emerge as the most frequently predicted commodity assets. In the realm of cryptocurrency forecasting, Bitcoin (BTC) and Ether (ETH) are the predominant focus. In contrast, forex rate forecasting receives relatively limited attention in the literature. Nonetheless, it is evident that all of these forecasting topics persistently underscore the importance of price forecasting.

Figure 6 summarises the characteristics of the raw datasets, including data type, length, frequency, and data source. Regarding data type and frequency, most included studies used daily historical trading data, indicating the easy accessibility and reliability of this type of data for research purposes. For dataset length, a large raw dataset containing more than five years of data is typically required to ensure sufficient model training. Yahoo Finance is the most widely used raw data source because it offers an accessible database for queries and downloads. Additionally, sources such as Wind, Tushare, and Joinquant specialise in offering financial market data in China, while sources such as the U.S. Energy Information Administration and World Gold Council provide access to energy market data and gold price data.

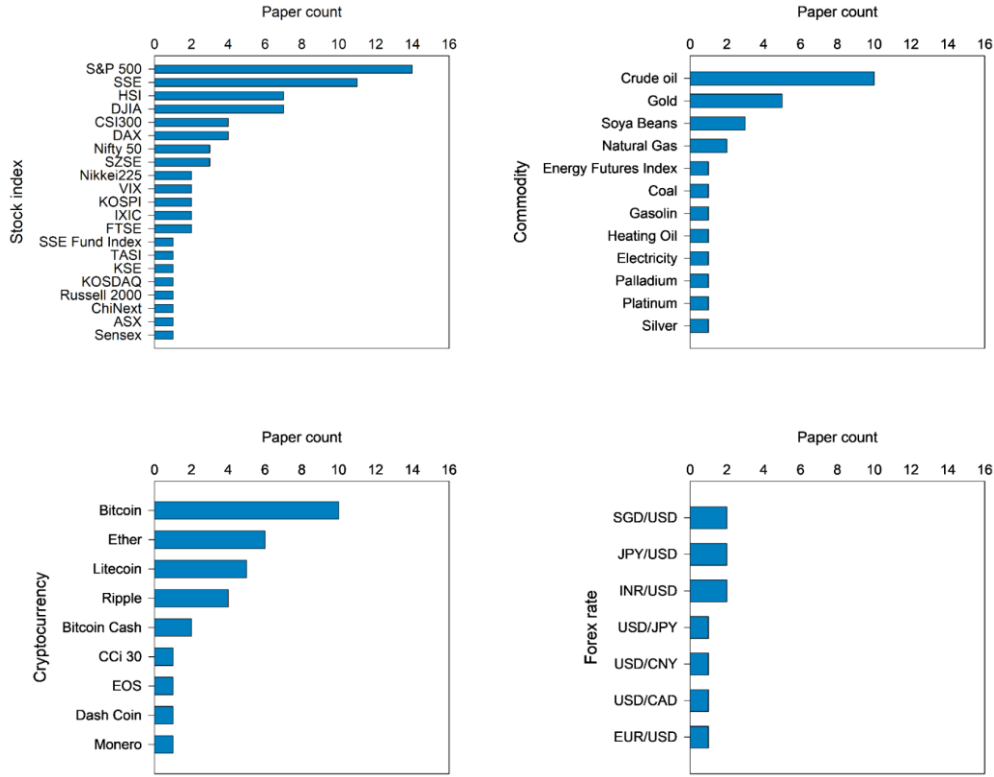

**Figure 5.** Summary of specific financial assets predicted in reviewed studies.

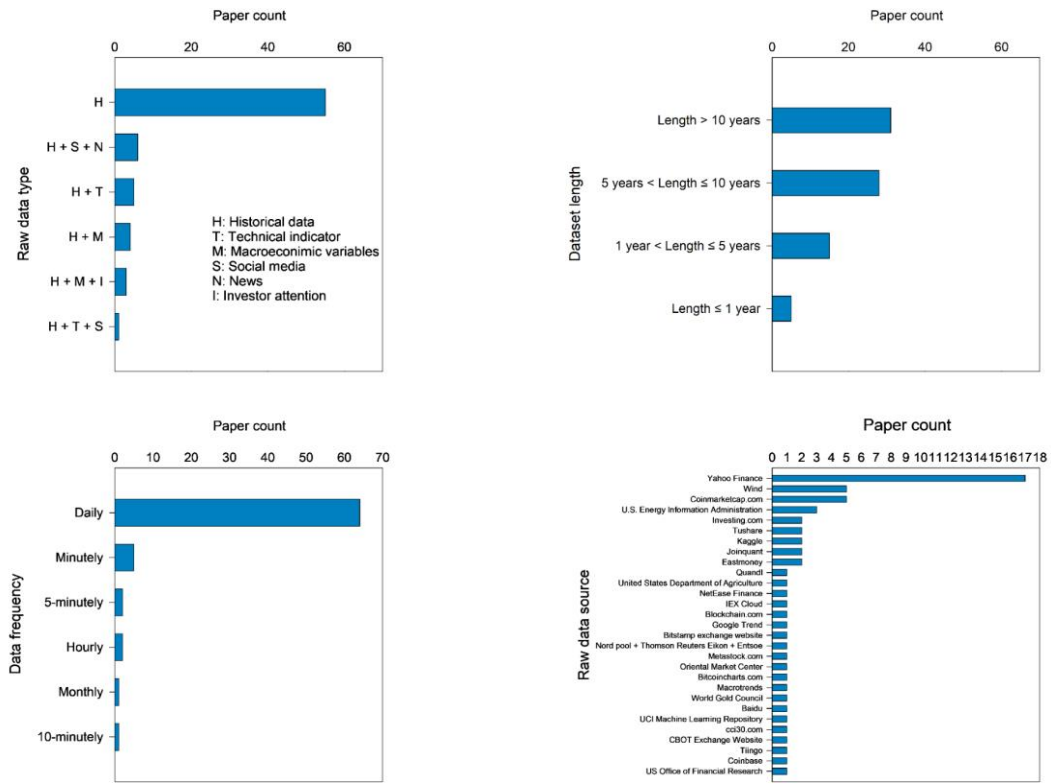

**Figure 6.** Summary of raw data characteristics: type, length, frequency, and source.

# 4. DEEP LEARNING MODELS FOR PRICE FORECASTING

In this section, we introduce different deep learning models proposed for price forecasting in the reviewed studies. These models are classified into two categories, namely, individual and ensemble models, based on their structure and nature. Individual models provide independent and complete forecasts, whereas ensemble models combine parallel models to generate collective predictions.

**4.1 Individual models**

*4.1.1 Deep neural networks (DNNs)*

A DNN is a collection of neurons organised in a sequence of multiple layers (Montavon et al., 2018). Figure 7 shows the architecture of the DNN for price forecasting. In contrast to a more primitive ANN with only one hidden layer, the DNN has multiple hidden layers. In a DNN model, neurons in each layer receive neuron activation from the previous layer and are connected to neurons in the subsequent layer to form a network structure. Each neuron performs simple computations, such as the weighted sum of the input, and activation functions are applied to the output of each neuron, thereby introducing nonlinearity and enabling complex computations. The connections between neurons are associated with weights adjusted using error backpropagation (Rumelhart et al., 1986).

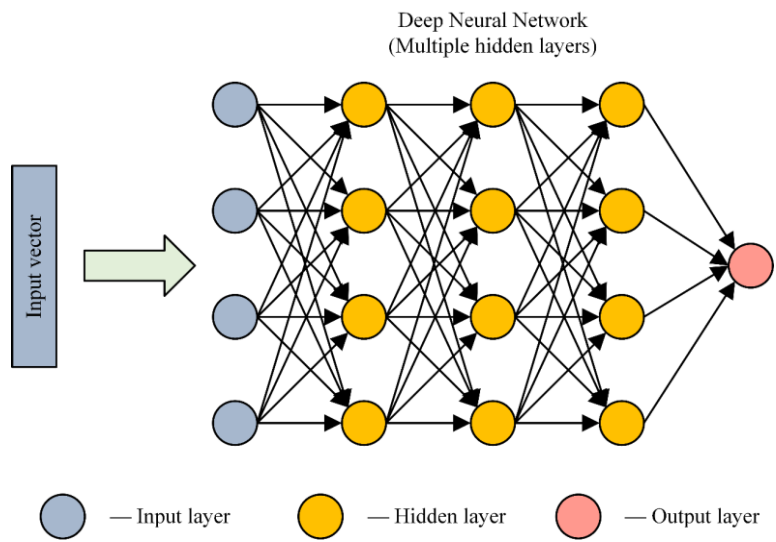

**Figure 7.** DNN for price forecasting.

According to Tripathi and Sharma (2022), DNNs demonstrate superior performance compared with LSTM and CNN-LSTM models when predicting BTC prices using technical indicators as model inputs. In addition, Omar et al. (2022) employed a DNN to process denoised inputs, and the proposed model outperformed traditional statistical methods such as ARIMA.

Compared with ANNs, which have limited representational power, DNNs can learn hierarchical representations of data, where lower layers capture simple features and higher layers combine them to form more complex abstractions. This ability enables them to learn more intricate and abstract features, leading to potentially improved model performance on complex tasks. However, conventional DNNs are not specifically designed for sequential data. They processed the input sequence as a flattened vector without explicitly considering the relationships between different elements in the original data. This limitation may affect the ability of the network to capture spatial or temporal dependencies between data points in sequential data, thereby constraining the applicability of DNNs as models for price forecasting.

*4.1.2 One-dimensional convolutional neural networks (1D CNNs)*

Inspired by the effectiveness of convolutional operations in feature extraction for image recognition, Waibel et al. (1989) designed a 1D CNN that performed convolutional operations on 1D sequences. The structure of the 1D CNN for price forecasting is illustrated in Figure 8. The core concept behind 1D CNNs is using one-dimensional convolutional filters that slide over the input sequence to capture local patterns or features at different positions (Li et al., 2022). Filters, also known as kernels, are the key components of the convolutional layer and perform convolutions by computing the weighted sums of the neighbouring elements. Activation functions, such as rectified linear units (ReLUs), are often applied to introduce nonlinearity. Additional layers such as pooling layers are often incorporated for downsampling.

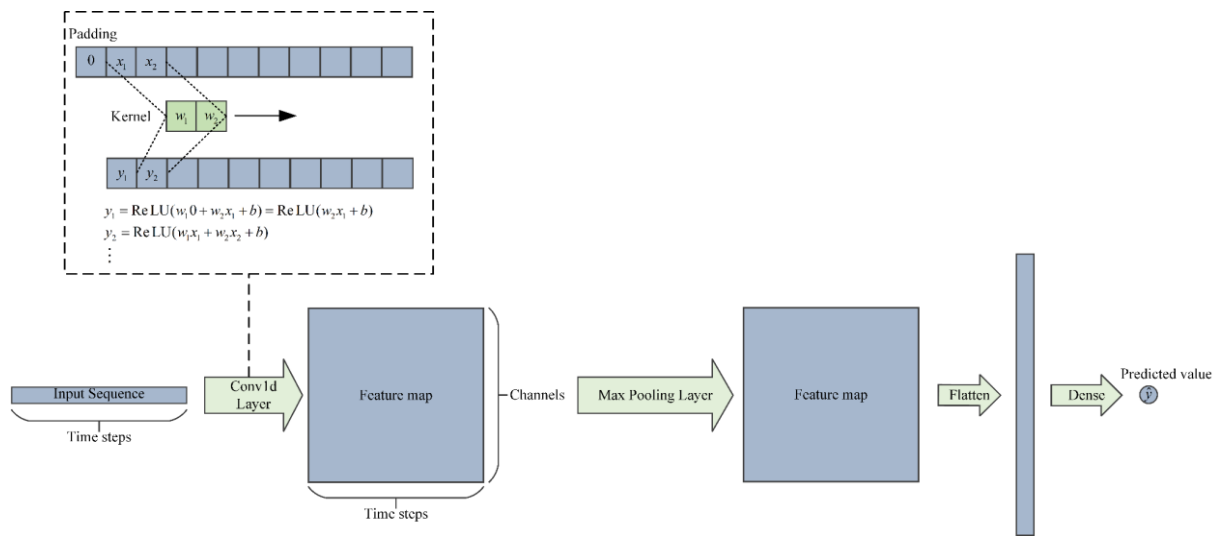

**Figure 8.** 1D CNN for price forecasting.

Durairaj and Mohan (2022) proposed a CNN-based model for price forecasting that incorporated chaos theory, 1D CNN, and polynomial regression. In their method, the financial time series underwent examination and modelling, considering the presence of chaos, before being fed into the 1D CNN for initial predictions.

Subsequently, polynomial regression provided an error prediction based on the error series obtained from the initial predictions. The error predictions were then added to the initial predictions to obtain the final predictions. 1D CNNs can learn hierarchical representations by stacking multiple convolutional layers, where lower layers capture simple local features and higher layers capture more complex global patterns. However, 1D CNNs may struggle to model long-term dependencies and relationships between distant data points in sequential data. Consequently, they have limited application in price forecasting.

*4.1.3 Recurrent neural networks (RNNs)*

RNNs are a type of neural network specifically designed to process sequential data (Rumelhart et al., 1985). An RNN model has a recurrent connection that enables information to flow from the previous steps to the current step, thereby capturing the temporal dependencies between data points at different positions in the sequence. However, simple RNNs suffer from the "vanishing gradient" problem, in which the influence of past inputs diminishes exponentially over time. Consequently, several variants of RNN cells have been developed to address the limitations of simple RNNs, as shown in Figure 9.

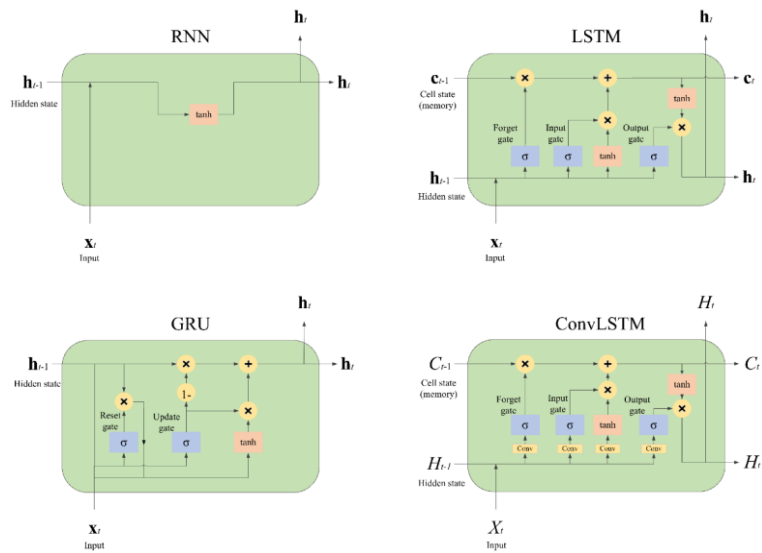

**Figure 9.** RNN cell and its variants. Based on (Rumelhart et al., 1985; Hochreiter & Schmidhuber, 1997; Cho, 2014; Shi et al., 2015).

The most well-known RNN cell variant is the LSTM cell, which was developed by incorporating a memory mechanism (Hochreiter & Schmidhuber, 1997). An LSTM cell can store information for long durations, selectively forget irrelevant information, and update its content based on new input. It consists of three primary components: an input gate, forget gate, and output gate. These gates play a crucial role in regulating the flow of information into, out of, and within the memory cell, thereby allowing LSTM networks to capture the long-term dependencies present in sequential data effectively. Another RNN cell variant, the gated recurrent unit (GRU), shares similarities with the LSTM cell but possesses a simplified structure (Cho, 2014). A GRU cell combines the input gate and forget gate into a single update gate and combines the cell state and hidden state into a single hidden state. It has fewer parameters than an LSTM cell, which makes it more computationally efficient. In addition, the convolutional LSTM (ConvLSTM) cell is an extension of the LSTM cell that incorporates convolutional operations into the LSTM structure (Shi et al., 2015). It has the same memory structure as the LSTM cell but replaces the fully connected operations with convolutional operations, enabling ConvLSTM networks to capture spatiotemporal dependencies in sequences of 2D matrices.

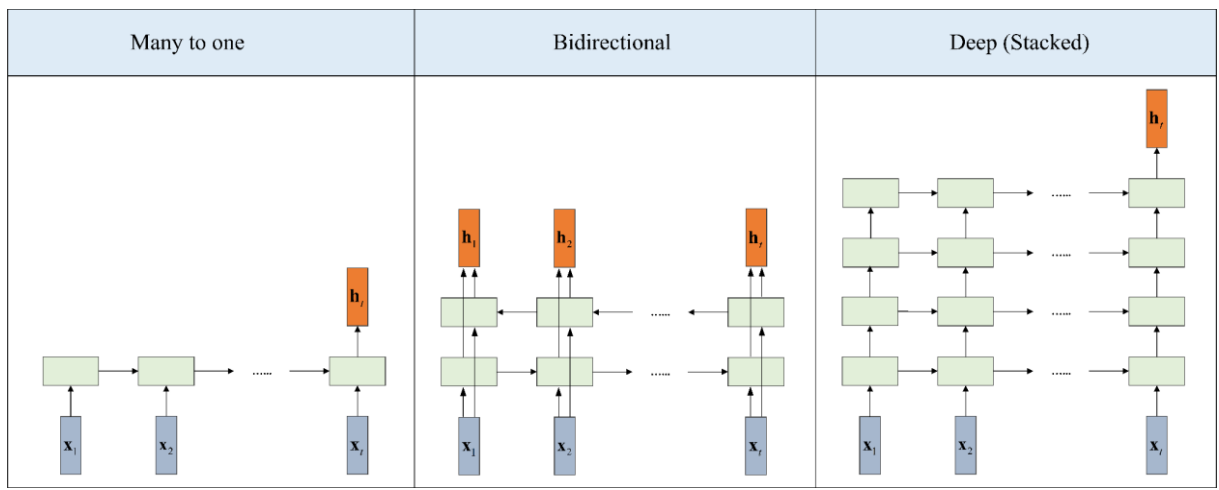

**Figure 10.** Different types of recurrent layers for price forecasting.

Based on these RNN cells, approximately three types of recurrent layers can be applied to price forecasting, as shown in Figure 10. The first type is the "many-to-one" recurrent layer, which uses a sequence of historical data as input and generates a 1D vector at the last time step. This 1D vector is passed through a dense layer with a suitable activation function to obtain the predicted value. The second type is the bidirectional recurrent layer, which is specifically designed to capture information from both past and future contexts (Schuster & Paliwal, 1997). It concurrently processes the input sequence in both the forward and backward directions, and subsequently merges the outputs of the two opposing layers. By incorporating knowledge of the future context, a bidirectional RNN can gain a deeper understanding of the input sequence, thereby enabling more informed predictions. The last type is the stacked recurrent layer, which consists of multiple recurrent layers stacked on top of each other (Pascanu et al., 2014). Each layer in the stack used the output sequence from the previous layer as its input. This layered architecture enables the network to capture hierarchical representations and to learn more intricate patterns within the input sequence.

Table 1 summarises the RNN-based forecasting models proposed in the reviewed studies. The ability of the LSTM network to capture temporal dependencies and patterns without suffering from the vanishing gradient problem has made it the most widely applied deep learning model for price forecasting. However, LSTM-based models have a limited memory capacity and may struggle to differentiate between intricate patterns and random fluctuations. Under these conditions, denoising methods, such as filtering techniques or smoothing algorithms, are often used to preprocess the data and remove or reduce the impact of noise before feeding the data into an LSTM model (Bukhari et al., 2020; Qiu et al., 2020). Moreover, incorporating domain knowledge and performing feature engineering can enhance model performance (Boubaker et al., 2022; Ji et al., 2021; Jin et al., 2020; Chen et al., 2020; Vuong et al., 2022). Appropriate feature engineering can help to expose relevant patterns or relationships that are not readily apparent in raw data and enable an LSTM-based model to learn more effectively.

**Table 1.** Summary of RNN-based models for price forecasting.

| Authors (year) | Datasets | Period | Feature set | Lag | Horizon | DL model | Baselines | Performance metrics |
|---|---|---|---|---|---|---|---|---|
| *Simple RNN-based models:* | | | | | | | | |
| Livieris et al. (2022) | Litecoin, Bitcoin, Ether, Ripple, CCi30 | 2017-2019 | C | 7 days, 14 days, 21 days | 1 day | Dropout weight-constrained RNN | SVR, LSTM, BiLSTM, CNN, KNN, Linear Regression | RMSE, MAE |
| Dixon and London (2021) | Bitcoin | 2018 | – | 4 mins | 4 mins | AlphaRNN | LSTM, RNN, GRU | MSE |
| Boubaker et al. (2022) | Crude oil | 1993-2021 | OHLCV+T | – | 4 weeks | CP-ADARNN | LSTM, GRU, ARIMA, Random Forest, Random walk, Lasso, ENet, Ridge | RMSE, MAE, Information coefficient, $R^2$ |
| *LSTM-based models:* | | | | | | | | |
| Chou et al. (2021) | Individual stocks | 2020 | OHLCV+S+E | – | 1 day | LSTM-AM | LSTM | RMSE, MAE, MAPE, $R^2$ |
| Yang et al. (2020) | Crude oil, Gasoline, Coal | 2009-2019 | OHLC | 4 days | 1 day | LSTMRT | LSTM, SVR, BPNN, EEMD-LSTM, VMD-LSTM | RMSE, MAE, MAPE, SMAPE, TIC |
| Tang et al. (2021) | DJIA | 2020 | C | – | 6 hours | WT-LSTM, SSA-LSTM | LSTM, RNN | RMSE, MAE, MAPE, SDAPE |
| Qiu et al. (2020) | S&P 500, HSI, DJIA | 2000-2019 | OHLCV | – | 1 day | WT-LSTM-AM | LSTM, WT-LSTM, GRU | MSE, RMSE, MAE, $R^2$ |
| Zhang et al. (2022) | Gold, Bitcoin | 2016-2021 | C | 50 days | 5 days | WT-LSTM-P | LSTM | MSE, RMSE, MAPE, $R^2$ |
| Ji et al. (2021) | Individual stocks | 2010-2019 | OHLCV+T+S | 7 days | 1 day | Doc-WT-LSTM | LSTM, RNN, ARIMA | RMSE, MAE, $R^2$ |
| Song et al. (2021) | SSE, S&P 500, KOSPI | 2001-2020 | C | 20 days | 1 day | P-FTD-LSTM | LSTM, RNN, P-FTD-RNN, GRU, P-FTD-GRU | RMSE, MAE, MAPE |
| Bukhari et al. (2020) | Individual stocks | 2009-2018 | O | 30 days | 1 day | ARFIMA-LSTM | GRNN, ARIMA, ARFIMA | RMSE, MAE, MAPE |
| Jin et al. (2020) | Individual stocks | 2013-2018 | OHLCV+S+E | – | 1 day | S-EMD-LSTM-AM | LSTM, LSTM-AM, Random Forest | RMSE, MAE, MAPE, $R^2$ |
| Chen et al. (2020) | S&P 500, Nasdaq, DJIA, Russell 200 | 2008-2019 | C+T | 10 days | 5 days | PCA-MLP-BiLSTM | SVR, LSTM, MLP, CNN, MLP-BiLSTM | MSE, MAE, EVS, MSLE, MedAE, $R^2$ |
| Chen and Zhou (2021) | Individual stocks | 2010-2020 | OHLV+T | 5 days | 1 day | GA-LSTM | PCA-SVR, DA-RNN, LSTM, Random Forest | MSE |
| Vuong et al. (2022) | EUR/USD | 2008-2018 | OHLCV+T | – | 5 mins | XGBoost-LSTM | ARIMA | MSE, MAE, RMSE |
| Song et al. (2020) | KOSPI, KOSDAQ | 2001-2019 | OHLCV | 20 days | 1 day | BIRCH-LSTM | LSTM, RNN | RMSE, MAE, MAPE |

| Authors (year) | Datasets | Period | Feature set | Lag | Horizon | DL model | Baselines | Performance metrics |
|---|---|---|---|---|---|---|---|---|
| G. Kumar et al. (2022) | S&P 500, Nifty 50, Sensex | 2015-2021 | OHLCV+T | – | 12 weeks | Adaptive PSO-LSTM | PSO-LSTM, GA-LSTM, LSTM, ENN | MSE, RMSE, SMAPE, TIC, MAAPE |
| R. Kumar et al. (2022) | Individual stocks | 2010-2020 | OHLCV+T+S | 9 days | 1 day | ABC-LSTM | DE-LSTM, GA-LSTM, LSTM | RMSE, MAPE |
| Durairaj and Mohan (2021) | Gold, Crude oil, Soya beans, SSE Composite, S&P 500, Nifty 50, INR/USD, JPY/USD, SGD/USD | 2014-2020 | C | – | 1 day | Chaos+LSTM+PR | LSTM, Chaos+LSTM, ARIMA, Prophet | MSE, TIC, D-Stat |
| Guo et al. (2021) | Bitcoin | 2015-2020 | OHLCV+S+E | 5 days | 1 day | MRC-LSTM | LSTM, CNN, CNN-LSTM, MLP | RMSE, MAE, MAPE, $R^2$ |
| Ko and Chang (2021) | Individual stocks | 2015-2020 | OHLCV+S+E | 20 days | 1 day | BERT+LSTM | LSTM | RMSE |
| *GRU-based models:* | | | | | | | | |
| B. Wang and Wang (2021) | Crude oil, Natural gas, Heating oil | 2012-2019 | C | – | 1 day | DBGRUNN | SVR, LSTM, ERNN, GRU, DBGRUNN, RIF-GRUNN | RMSE, MAE, SMAPE, TIC, $R^2$ |
| Li et al. (2020) | Soya Beans | 1956-2020 | C+M+T | 299 days | 1 day | DP-MAELS: VAE+GRU+AR | SVM-VAR, LSTM, CNN, RNN, ARIMA | MAPE, TIC, Correlation coefficient, RSE |
| Li and Wang (2020) | Energy future index | 2009-2019 | C | – | 1 day | ST-GRU | LSTM, BPNN, EEMD-SVR, GRU, WNN, WNNRT | RMSE, MAE, SMAPE, TIC, Directional symmetry, Correct uptrend, Correct downtrend, Correlation coefficient, MCCS |
| *Stacked RNN-based models:* | | | | | | | | |
| T. Niu et al. (2020) | SSE, DJIA, SSE Fund Index | 1998-2015 | OHLCV+T | 100 days | 1 day | ConvLSTM-LSTM-GRU | MLP, ARIMA, GPR, WNN, ConvLSTM-LSTM, ConvLSTM-GRU, ConvLSTM-LSTM-GRU | RMSE, MAE, MAPE, TIC, MdAPE |
| Wang et al. (2020) | CSI300 | 2017-2019 | OHLCV+S+E | – | 1 day | ConvLSTM-ConvLSTM | LSTM+Attention, SVR, LSTM, Linear Regression, ConvLSTM-AM, Seq2Seq | MSE, TIC, Directional symmetry |
| Yu and Yan (2020) | S&P 500, HSI, DJIA, CSI300, Nikkei225, ChiNext | 2008-2017 | C | 3 days | 1 day | PSR-Stacked-BiLSTM | SVR, MLP, ARIMA | RMSE, MAPE, Correlation coefficient |
| Nasirtafreshi (2022) | Litecoin, Bitcoin, Ether, Bitcoin Cash | 2016-2018 | C | 3 days | 1 day | RNN-LSTM | SVR, LSTM, MLP, CNN, GRU, LSTM-GRU, ARIMA, Random Forest, GRU-CNN, XGBoost | RMSE, MAE, MAPE, $R^2$ |

Feature set: Open (O), High (H), Low (L), Close (C), Volume (V), Technical indicator (T), Sentiment indicator (S), Event features (E), Macroeconomic indicator (M).

*4.1.4 Convolutional-recurrent neural networks (CRNNs)*

CRNNs were a type of hybrid neural network developed by combining the complementary strengths of CNNs and RNNs to capture spatiotemporal patterns within the data in various tasks, including time series forecasting (Shi et al., 2017; Zhan et al., 2018). Figure 11 illustrates the typical architecture of a CRNN for price forecasting. In this architecture, a 1D CNN extracts spatial features from the input sequence. The output of the 1D CNN, which is a sequence of feature vectors, is then fed into a recurrent layer that captures temporal dependencies and long-term patterns in the feature map. It is worth noting that the recurrent layers can be placed before the convolutional layers when temporal dependencies within the input sequence are considered crucial and must be modelled before extracting spatial features. (Kanwal et al., 2022; Zhang et al., 2021).

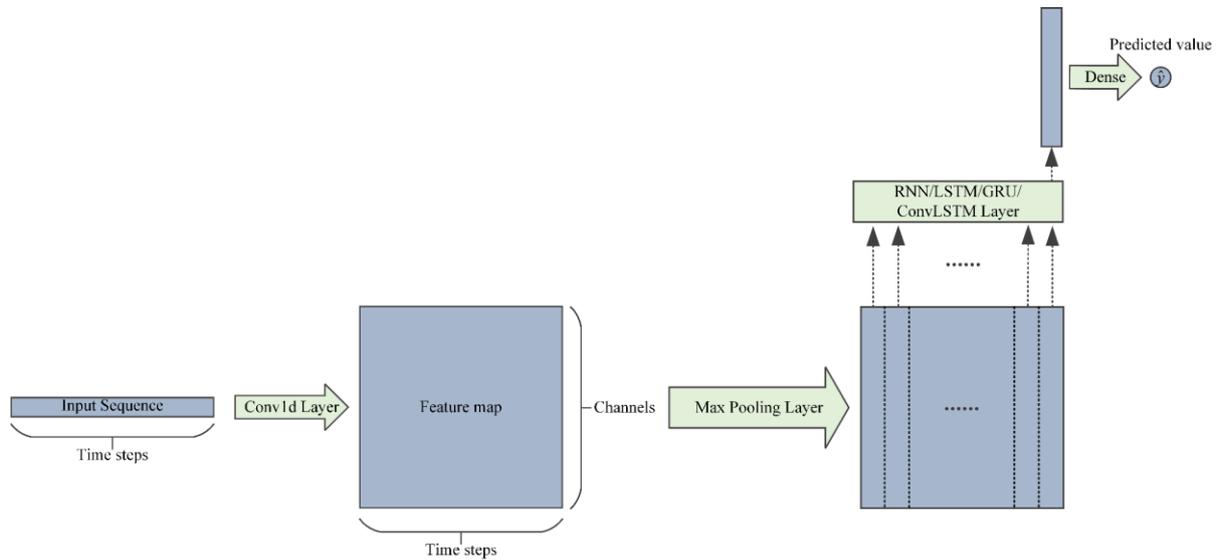

**Figure 11.** CRNN for price forecasting.

Table 2 summarises the CRNN-based forecasting models proposed in the reviewed studies. The advantages of the LSTM layers in handling long-term dependencies make them the preferred choice for constructing CRNN architectures. Although CRNNs are suitable for processing sequential data with spatiotemporal dynamics, they have certain limitations. CRNNs are computationally expensive and require significant resources owing to the combination of convolutional and recurrent layers, resulting in an increased training time. The complexity of CRNNs can make hyperparameter tuning more challenging than that for standalone CNN or RNN models. In addition, the intertwining of the spatial features extracted by the CNN and temporal dependencies captured by the recurrent layers can make the interpretation and understanding of CRNNs more difficult. Despite their effectiveness in capturing spatiotemporal patterns, careful consideration of computational resources, training time, and interpretability is necessary when utilising CRNNs in practical applications.

**Table 2** Summary of CRNN-based models for price forecasting.

| Authors (year) | Datasets | Period | Feature set | Lag | Horizon | DL model | Baselines | Performance metrics |
|---|---|---|---|---|---|---|---|---|
| Livieris et al. (2020b) | Gold | 2014-2018 | C | 9 days | 1 day | CNN-LSTM | SVR, FFNN, LSTM | RMSE, MAE |
| Lu et al. (2020) | SSE Composite | 1991-2020 | OHLCV+T | 10 days | 1 day | CNN-LSTM | LSTM, MLP, CNN, RNN, CNN-RNN | RMSE, MAE, $R^2$ |
| Livieris et al. (2020a) | Natural gas | 2015-2018 | C | – | 1 day | CNN-LSTM | SVR, ANN, DTR | RMSE, MAE |
| Aldhyani and Alzahrani (2022) | Individual stocks | 2014-2017 | C | – | 1 day | CNN-LSTM | LSTM | MSE, RMSE, NRMSE, $R^2$ |
| H. Wang et al. (2021) | SZSE Composite | 1991-2020 | OHLCV+T | 5 days | 1 day | CNN-BiLSTM | LSTM, BiLSTM, MLP, RNN, CNN-LSTM, CNN-BiLSTM | RMSE, MAE, $R^2$ |
| Zheng (2021) | ETF | 2018-2020 | OHLCV+T | 4 days | 1 day | CNN-BiLSTM-AM | LSTM, ARIMA, DNN | MSE, MAPE, $R^2$ |

| Authors (year) | Datasets | Period | Feature set | Lag | Horizon | DL model | Baselines | Performance metrics |
|---|---|---|---|---|---|---|---|---|
| Chen et al. (2021) | Individual stocks, SSE Composite, CSI300 | 1990-2020, 2002-2021, 2002-2021 | OHLC+T | 10 days | 1 day | CNN-BiLSTM-ECA | LSTM, CNN-LSTM-ECA, BiLSTM, BiLSTM-ECA, CNN-LSTM, CNN-BiLSTM, CNN | MSE, RMSE, MAE |
| Jaiswal and Singh (2022) | Individual stocks | 2000-2021 | – | – | 1 day | CNN-GRU | CNN-RNN, CNN-LSTM | RMSE, $R^2$ |
| Kang et al. (2022) | Bitcoin, Ether, Ripple | 2021 | C | – | 1 min | CNN-GRU | LSTM, CNN-LSTM, BiLSTM, RNN, ARIMA, Prophet, XGBoost | RMSE |

Feature set: Open (O), High (H), Low (L), Close (C), Volume (V), Technical indicator (T).

*4.1.5 Autoencoders*

Autoencoders are deep learning models that learn efficient data representations using an encoder-decoder framework (Baldi, 2012). For time series forecasting, this type of model leverages the encoder to capture compressed representations of historical time series data and the decoder to generate predictions for future time steps. Figure 12 shows the architecture of the LSTM-based autoencoder for price forecasting. In applications, there are multiple choices when constructing an encoder and a decoder. Choudhury et al. (2020) proposed an autoencoder for stock price forecasting using LSTM networks as both the encoder and decoder. In contrast, Li and Becker (2021) proposed three types of encoders for autoencoder models, including CNN, LSTM, and ConvLSTM networks. Iqbal et al. (2021) proposed an autoencoder architecture in which the encoder was a ConvLSTM network.

Using autoencoders for price forecasting offers the advantage of feature learning, which enables the capture of relevant patterns and nonlinear relationships in data. However, the drawbacks of using autoencoders include the lack of interpretability, the need for sufficient training data, and the risk of overfitting.

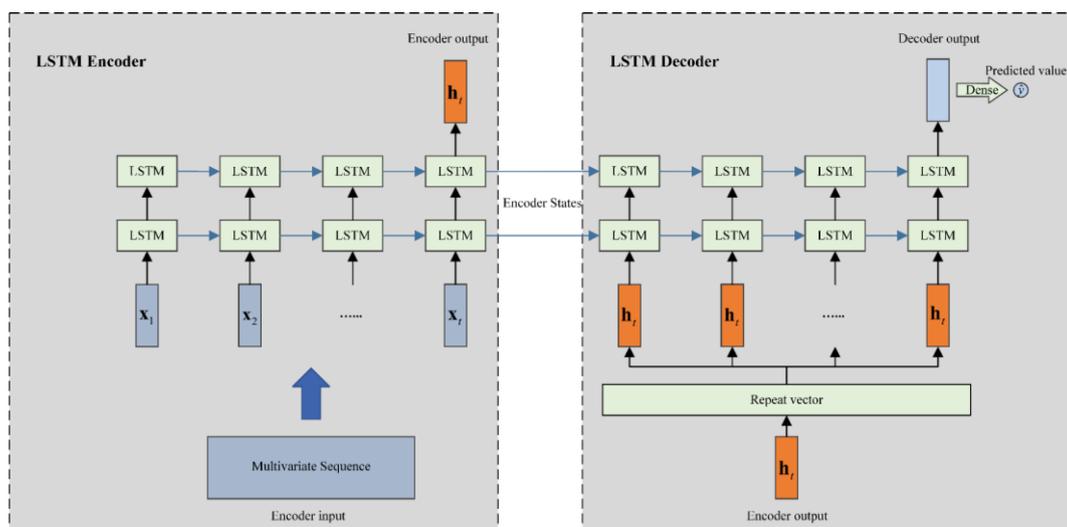

**Figure 12.** LSTM-based autoencoder for price forecasting.

*4.1.6 Transformers*

The development of the Transformer model stems from the need to effectively model sequential data with long-range dependencies, as observed in tasks such as natural language processing (NLP) (Vaswani et al., 2017). The ability to process long sequential data makes the Transformer model a suitable option for price forecasting. Figure 13 shows the price forecasting model based on a Video Transformer (ViT) (Malibari et al., 2021). In this architecture, the multivariate sequence is divided into fixed-length nonoverlapping patches. These patches are linearly projected onto lower-dimensional embeddings. The embedded patches are then fed to a standard Transformer encoder to generate the predicted value.

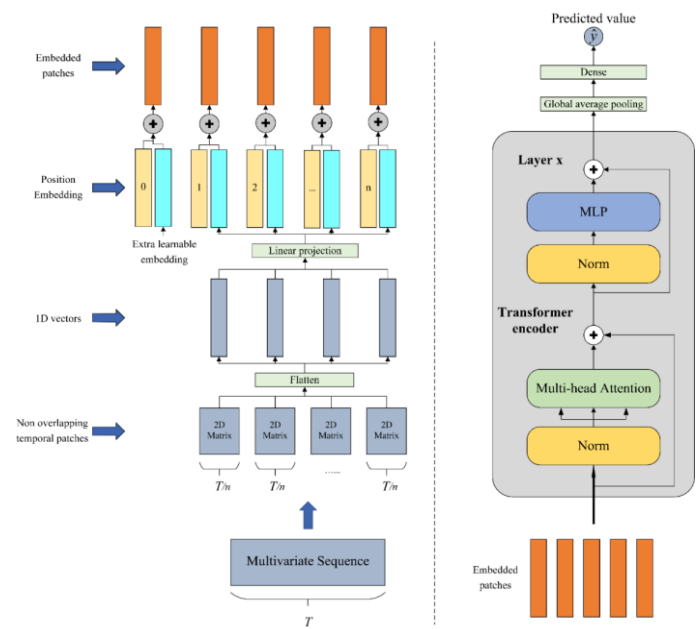

**Figure 13.** Video Transformer for price forecasting. Based on (Malibari et al., 2021).

As depicted in Figure 14, the key component of a Transformer is the multi-head attention, which leverages several self-attention layers to jointly attend to different parts of the input sequence multiple times, enabling it to capture diverse and fine-grained relationships between distant data points. Its input comprises three identical matrices, namely the query (Q), key (K), and value (V) representations. After linear transformation, the attention scores between the query and key representations are calculated and subsequently combined with the value representation using weighted averages to produce the final output representation. This process can be conducted multiple times in parallel, and the resulting attention outputs are concatenated and linearly transformed to obtain an output that is subsequently used for downstream tasks.

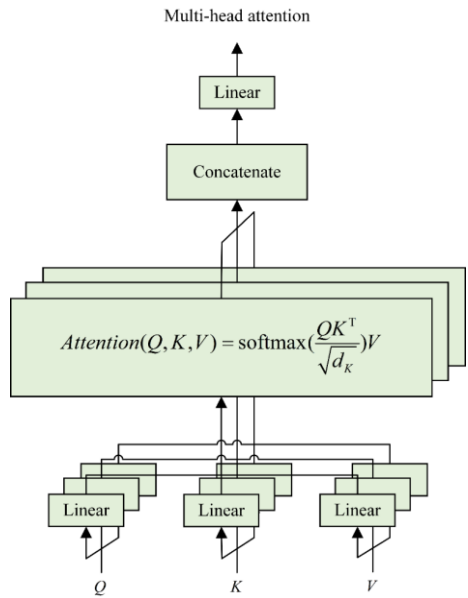

**Figure 14**. Multi-head attention. Based on (Vaswani et al., 2017).

Transformers utilise self-attention mechanisms when dealing with long sequences, allowing them to capture dependencies between distant positions in a sequence more effectively than LSTM networks. One potential drawback of the Transformer model is its computational complexity because the self-attention mechanism requires quadratic time and memory with respect to the input sequence length, making it less efficient for extremely long sequences.

*4.1.7 Graph neural networks (GNNs)*

GNNs are a class of deep learning models designed to operate on graph-structured data, allowing for effective representation learning and reasoning over nodes and edges in the graph (Scarselli et al., 2009). Figure 15 shows the GNN model for price forecasting (Patil et al., 2020). The model input is a three-dimensional tensor generated based on a graph constructed using a network of related assets. Each node in this graph represents one asset, and the edges are established based on asset relationships such as correlation or sector similarity. The feature set of each node includes the attributes of the historical trading data and other technical indicators. Node feature matrices are stacked along the temporal dimension to construct the model input.

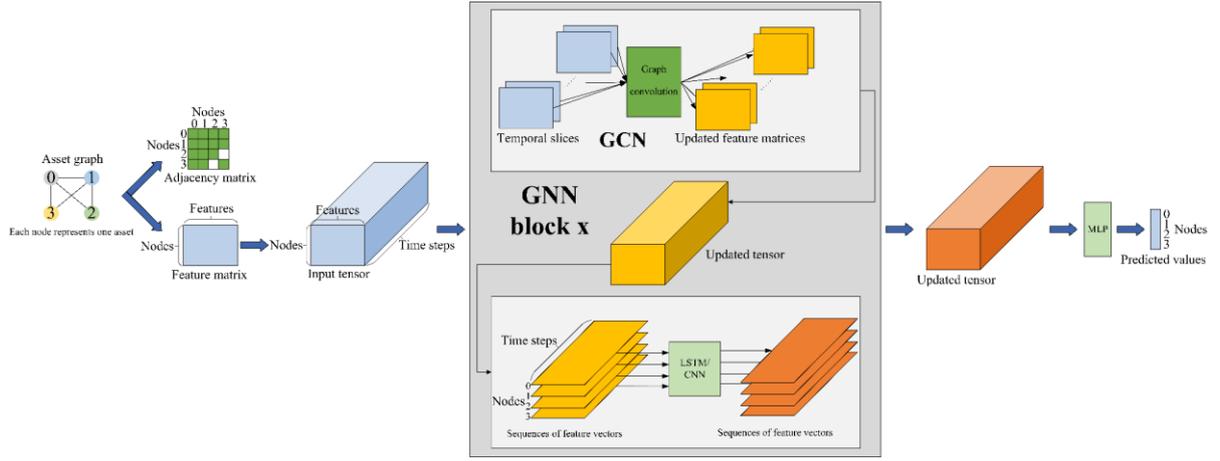

**Figure 15.** GNNs for price forecasting. Based on (Patil et al., 2020).

The key component of a GNN model is the GNN block, which includes a graph convolution network (GCN) and a CNN or LSTM network. The CNN or LSTM network can be placed before the GCN to first perform feature extraction (Yin et al., 2022). The GCN leverages the adjacency matrix to determine the node relationships and update the node features. An adjacency matrix is regarded as a fixed representation of a graph structure. It is defined by the connections or relationships between the nodes in the graph, represents the static structural information of the graph and remains constant throughout the GCN computation. When the adjacency matrix is used as the kernel for graph convolution, the equation for node update can be defined as follows:

$$\mathbf{H}' = \sigma(\mathbf{AHW}), \qquad (2)$$

where $\mathbf{H}$ denotes the input feature matrix with shape $(N, D)$. $N$ is the number of nodes in the graph, and $D$ is the number of feature attributes per node before graph convolution. $\mathbf{H}'$ is the output feature matrix with shape $(N, F)$. $F$ is the number of feature attributes per node after graph convolution, and $\mathbf{A}$ is the adjacency matrix of the graph with shape $(N, N)$. Each entry $\mathbf{A}[i, j]$ represents the strength or presence of an edge between nodes $i$ and $j$. $\mathbf{W}$ represents the weight matrix with shape $(D, F)$ used in the graph convolutional operation. $\sigma$ is the activation function applied elementwise to obtain the output features $\mathbf{H}'$. After graph convolution, all the feature matrices are updated and used for downstream tasks.

By incorporating a graph structure, GNNs can model the interconnections among different assets and leverage information from related assets, potentially leading to enhanced forecasting performance. However, GNNs require careful construction of the input graph, including the determination of relevant relationships and selection of appropriate features, which can be challenging in financial markets. In addition, GNNs may struggle with limited historical data or sudden market shifts because the adjacency matrix relies heavily on past information. Finally, GNNs have high computational requirements, particularly for large-scale financial networks. Careful consideration of the data structure and model limitations is crucial for GNN application.

*4.1.8 Generative adversarial networks (GANs)*

GANs are deep learning models that consist of a generator and discriminator trained in a competitive setting (Goodfellow et al., 2014). The training process of the GANs is similar to a minimax game between the generator and discriminator, in which the generator aims to minimise the discriminator's ability to distinguish between real and generated samples, whereas the discriminator aims to maximise its discrimination capability. Theoretically, the training of the GAN model is considered to have been completed when the discriminator achieves a probability of 0.5 in discerning real samples from generated ones, which indicates the attainment of the Nash equilibrium in the minimax game. Although GANs are primarily used for generative tasks, they have also been applied to price forecasting, as shown in Figure 16. In a GNN model for price forecasting, the generator is typically an LSTM network or its variants, which is trained to generate realistic future prices, whereas the discriminator is usually a neural network for classification tasks, such as a CNN or a feed-forward neural network (Li et al., 2021; Staffini, 2022).

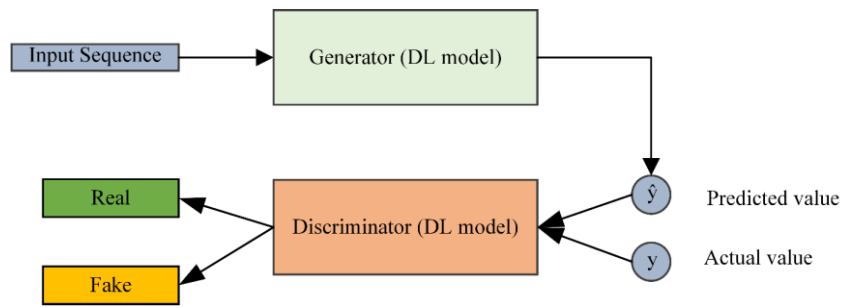

**Figure 16.** GANs for price forecasting. Based on (Staffini, 2022).

GANs can generate synthetic price sequences that closely resemble real market data. However, they require large amounts of data for effective training, which may be a challenge when the amount of data is limited. In addition, GANs are sensitive to training dynamics and may become unstable during training.

*4.1.9 Deep quantum neural networks (DQNNs)*

DQNNs are a type of neural network that combine concepts from quantum computing and classical deep learning (Beer et al., 2020). A DQNN model typically comprises a quantum neural network (QNN) designed for quantum prediction and a classical DNN that serves as a measurement apparatus. It leverages the principles of quantum mechanics, such as superposition and entanglement, to perform computations that could offer advantages over classical neural networks. Paquet and Soleymani (2022) proposed a DQNN model named QuantumLeap for price forecasting. This forecasting model assumes that several entangled "particles" are arranged in a fixed order and that the state of each particle is in a two-level Hilbert space, with all Hilbert spaces being distinguishable. The current joint quantum state of these particles, denoted as $\left|\psi_t^{in}\right\rangle$, is encoded using a multivariate sequence that consists of "high" and "low" prices of the related financial asset:

$$\left|\psi_t^{in}\right\rangle = \left|L_{t-\Delta+1} + iH_{t-\Delta+1}, L_{t-\Delta+2} + iH_{t-\Delta+2}, \cdots, L_t + iH_t\right\rangle, \quad (3)$$

where $t$ denotes the current time step and $\Delta$ denotes the number of entangled particles. $L$ and $H$ denote the "low" and "high" prices in the multivariate sequence, respectively. Suppose the forecast horizon is $\tau$; then, the future joint quantum state of these particles, denoted as $\left|\psi_{t+\tau}^{out}\right\rangle$, can be encoded as

$$\left|\psi_{t+\tau}^{out}\right\rangle = \left|L_{t-\Delta+\tau+1} + iH_{t-\Delta+\tau+1}, L_{t-\Delta+\tau+2} + iH_{t-\Delta+\tau+2}, \cdots, L_{t+\tau} + iH_{t+\tau}\right\rangle. \quad (4)$$

Subsequently, $\left|\psi_t^{in}\right\rangle$ and $\left|\psi_{t+\tau}^{out}\right\rangle$ are normalised, and the input and output density matrices are obtained as follows:

$$\rho_t^{in} = \left|\psi_t^{in}\right\rangle\left\langle\psi_t^{in}\right|, \rho_{t+\tau}^{out} = \left|\psi_{t+\tau}^{out}\right\rangle\left\langle\psi_{t+\tau}^{out}\right|, \quad (5)$$

where $\left\langle\psi_t^{in}\right|$ and $\left\langle\psi_{t+\tau}^{out}\right|$ denote the conjugate transposes of $\left|\psi_t^{in}\right\rangle$ and $\left|\psi_{t+\tau}^{out}\right\rangle$, respectively. Correspondingly, $\{(\rho_t^{in}, \rho_{t+\tau}^{out})\}$ constructs a training set for the QNN that predicts the output density matrix, and $\{(\rho_{t+\tau}^{out}, H_{t+\tau})\}$ constructs a training set for the classical DNN that measures the maximum price reached by the related asset during a trading day from the output density matrix. Each hidden layer of the QNN comprises a series of noncommuting unitary operators through which the density matrix evolves while preserving its fundamental properties, such as the normalisation and conservation of probability. The unitary operators are updated recursively layer-by-layer until the cost function of the QNN reaches its maximum value. This cost function is determined by fidelity, which is a distinctive measure of the proximity between the output of the QNN and the desired output averaged over the training data (Paquet & Soleymani, 2022).

Although QuantumLeap demonstrates accuracy in stock price forecasting, challenges remain in applying this type of model, such as the requirement for specialised programming packages and hardware. Moreover, the training and optimisation of QNNs can be difficult owing to the lack of classical optimisation algorithms that can be directly applied. Overall, although DQNNs hold significant promise, their evolution and integration into practical applications are currently in the early stages. Several technical and practical obstacles must be overcome to unlock their complete potential and establish a definitive advantage over classical models.

## 4.2. Ensemble models

*4.2.1 Stacking ensemble*

Stacking ensembles leverage a combination of diverse base models to improve the overall performance and robustness of predictions and reduce the impact of individual model biases or errors (Bishop & Nasrabadi, 2006). The structure of the stacking ensemble is illustrated in Figure 17. The predictions of each base model are typically combined using meta-learning or the arithmetic average.

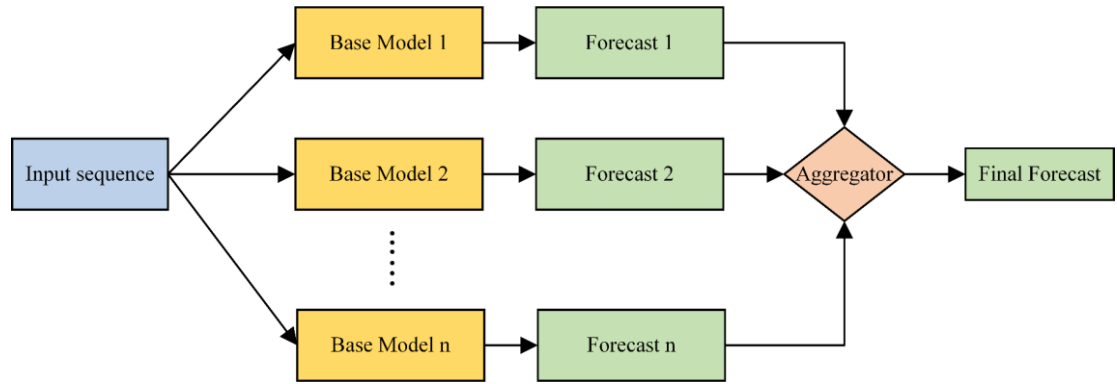

**Figure 17.** Stacking ensemble for price forecasting.

Table 3 summarises the stacking ensembles proposed in the reviewed studies. This type of ensemble can capture a wide range of patterns and effectively handle complex relationships in financial data, making them robust to different market conditions and more reliable than the individual models. However, the stacking ensembles require additional computational resources owing to increased model complexity. They may be sensitive to the selection and configuration of the base models. In addition, the interpretability of the stacking ensembles may be diminished because the combination of multiple individual models can make it challenging to understand the underlying factors contributing to the forecasts.

**Table 3.** Summary of stacking ensembles for price forecasting.

| Authors (year) | Datasets | Period | Feature set | Lag | Horizon | Base models | Aggregator | Baselines | Performance metrics |
|---|---|---|---|---|---|---|---|---|---|
| Patel et al. (2020) | Litecoin, Monero | 2016-2020 2015-2020 | C | 30 days | 1 day | LSTM, GRU | Dense layer | LSTM | MSE, RMSE, MAE |
| Livieris et al. (2021) | Bitcoin, Ether, Ripple | 2017-2020 | C | 7days 14 days | 1 day | CNN-LSTM | Concatenate layer | CNN-LSTM | RMSE, MAE, $R^2$ |
| J. Wang et al. (2021) | SSE Composite, S&P 500, SZSE Composite, DAX | 2011-2020 | C | 7 days | 1 day | BiLSTM, BiGRU | Weighted Sum | BiLSTM, BPNN, GRU, Random Forest | RMSE, MAE, MAPE |
| Lu and Peng (2021) | Individual stocks | 2010-2019 | OHLCV+T | 15 days | 1 day | MLP, LSTM, GRU, RNN | Stacking | LSTM, MLP, RNN, GRU | MSE, MAE, MAPE, $R^2$ |
| Yang et al. (2022) | Individual stocks | 2011-2020 | C+T | 4 days | 1 day | CNN, GRUA-FC | Dense layer | CNN, GRU, GRU-AM, Ensemble: (CNN+GRU) | RMSE, MAE, MAPE, $R^2$ |

| Authors (year) | Datasets | Period | Feature set | Lag | Horizon | Base models | Aggregator | Baselines | Performance metrics |
|---|---|---|---|---|---|---|---|---|---|
| Liu and Huang (2021) | Crude oil | 2007-2020 | OHLCV+S+E | 30 days | 1 day | ARIMA, LSTM, ARIMA-GARCH | Stacking | SVR, LSTM, ARIMA, ARIMA-GARCH | RMSE, MAPE, D-stat |
| Chong et al. (2020) | Individual stocks | 2004-2018 | OHLC | – | 1 day | CNN, LSTM, ConvLSTM | Average | CNN, Random Forest, RNN | RMSE |

Feature set: Open (O), High (H), Low (L), Close (C), Volume (V), Technical indicator (T), Sentiment indicator (S), Event features (E).

*4.2.2 Decomposition ensemble*

Decomposition ensembles combine decomposition techniques and deep learning models to enhance the predictive accuracy. As shown in Figure 18, in a decomposition ensemble, the original time series is first decomposed into a set of spectra known as intrinsic mode functions (IMFs). Subsequently, the sequences of each IMF are fed into an independent deep learning model for training. The predictions for each IMF are then summed to generate the final prediction of the target variable.

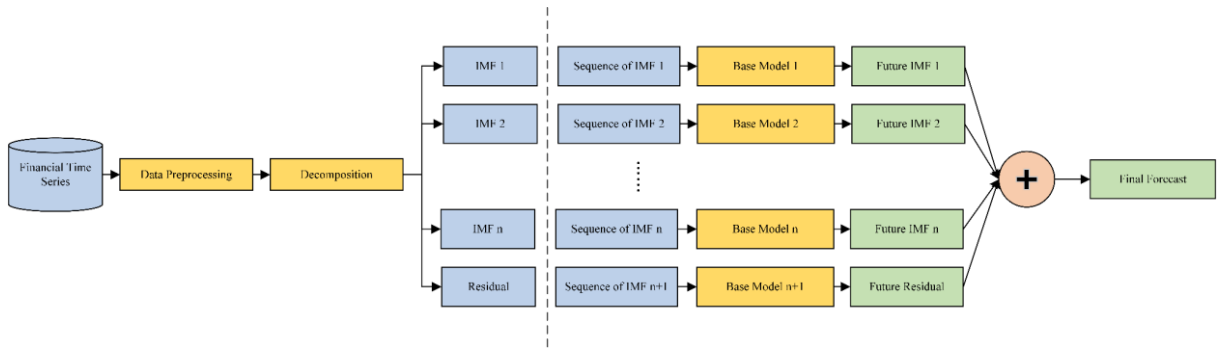

**Figure 18.** Decomposition ensemble for price forecasting.

Table 4 summarises the decomposition ensemble models proposed in the literature. Various signal decomposition techniques have been employed to perform time series decomposition, including empirical mode decomposition (EMD), ensemble empirical mode decomposition (EEMD), multivariate ensemble empirical mode decomposition (MEMD), and complete ensemble empirical mode decomposition (CEEMD).

**Table 4.** Summary of decomposition ensembles for price forecasting.

| Authors (year) | Datasets | Period | Feature set | Lag | Horizon | Decomposition | Base models | Baselines | Performance metrics |
|---|---|---|---|---|---|---|---|---|---|
| Shu and Gao (2020) | SSE Composite | 2001-2018 | C | 1000 days | 1 day | EMD | CNN-LSTM | CNN-LSTM, EMD-LSTM, EMD-SVR, SVR, LSTM, Persistence model | RMSE, MAE |
| Wang and Wang (2020) | Crude oil | 2009-2020 | C | – | 3 days | EMD | SW-GRU | EEMD-SW-GRU, EMD-BPNN, EMD-SVR, EMD-GRU | RMSE, MAPE, SMAPE, TIC |
| Xuan et al. (2020) | SSE Composite | 2015-2019 | OHLCV+T | – | 1 day | EMD | LSTM-CSI | LSTM-AM, EMD-LSTM, EMD-LSTM-AM, SVR, LSTM | RMSE, MAE, MAPE |
| Guo et al. (2022) | Individual stocks | 2015-2020 | – | – | 1 day | EEMD | Cluster-SVR-PSO-LSTM | PSO-LSTM, EEMD-PSO-LSTM, EEMD-SVR, LSTM, BPNN, ELM | MSE, RMSE, MAE, D-stat |
| Rezaei et al. (2021) | S&P 500, DJIA, DAX, Nikkei225 | 2010-2019 | C | 250 days | 1 day | CEEMD | CNN-LSTM | EMD-LSTM, EMD-CNN-LSTM, CEEMD-LSTM, LSTM, CNN-LSTM | RMSE, MAE, MAPE |

| Authors (year) | Datasets | Period | Feature set | Lag | Horizon | Decomposition | Base models | Baselines | Performance metrics |
|---|---|---|---|---|---|---|---|---|---|
| Lin and Sun (2020) | Crude oil | 1986-2019 | C | 32 days | 1 day | CEEMDAN | Stacked GRU | EEMD-Stacked GRU, EEMD-GRU, EEMD-LSTM, EEMD-LSSVR, EEMD-ANN, CEEMDAN-Stacked GRU, CEEMDAN-GRU, CEEMDAN-LSTM, CEEMDAN-LSSVR, CEEMDAN-ANN, LSSVR, LSTM, ANN, GRU, Staked GRU, ARIMA, Naïve | RMSE, MAPE |
| Deng et al. (2022) | SSE Composite, S&P 500, HSI | 2010-2020 | OHLCV | – | 10 days | MEMD | LSTM | LSTM, BPNN, EMD-LSTM, ARIMA | RMSE, MAPE, Directional symmetry |
| Lin et al. (2022) | Gold, Silver, Platinum, Palladium | 2009-2020 | C | – | 10 days | MEEMD | LSTM | SVR, LSTM, MLP, EEMD-SVR, EEMD-LSTM, EEMD-MLP, MEEMD-SVR, MEEMD-MLP | RMSE, MAE, MAPE, SMAPE |
| H. Niu et al. (2020) | HSI, FTSE, Nasdaq Composite | 2010-2019 | C | 4 days | 1 day | VMD | LSTM | EMD-BPNN, EMD-ELM, EMD-CNN, EMD-LSTM, VMD-BPNN, VMD-ELM, VMD-CNN, LSTM, CNN, BPNN, ELM | RMSE, MAE, MAPE |
| Yujun et al. (2021) | S&P 500, HSI, SZSE, DAX, VIX, ASX | Opening data-2020 | C | – | 1 day | VMD | LSTM | SVR, BARDR, RFR, KNR | RMSE, MAE, $R^2$ |
| Huang et al. (2021) | Crude oil, HSI, FTSE, VIX, USD/CAD, USD/CNY, USD/JPY | 2008-2013, 2010-2016, 2010-2017 | C | 31 days | 1 day | VMD | LSTM | EMD-SVR, EMD-Lahmiri, EMD-RKELM, EMD-SVNN, VMD-Lahmiri, VMD-SVNN, VMD-SVR, ARIMA, ARMA, SVNN, SVR, FFNN, LSTM, ELM | RMSE, MAE, MAPE |
| Liu and Long (2020) | Individual stocks, S&P 500, DJIA | 2010-2013, 2013-2016, 2014-2017 | C | 10 days | 1 day | EWT | dLSTM-PSO-ORELM | LSTM, PSO-LSTM, PSO-LSTM-ORELM, PSO-LSTM-RELM, PSO-LSTM-ELM, BPNN, EWT-PSO-LSTM | RMSE, MAE, MAPE, SMAPE, TIC |
| Althelaya et al. (2021) | S&P 500 | 2010-2018 | C | 5 days, 10 days | 1 day, 1 month | EWT | Stacked LSTM | Stacked LSTM | RMSE, MAE, MAPE, $R^2$ |
| J. Wang and Wang (2021) | Crude oil | 2009-2019 | C | 4 days | 1 day | WPD | SW-LSTM | SVR, LSTM, BPNN, WPD-BPNN, WPD-LSTM | RMSE, MAE, MAPE, SMAPE, TIC |

Feature set: Open (O), High (H), Low (L), Close (C), Volume (V).

Decomposition techniques effectively capture time series components (e.g., trends, seasonality, and residuals), thereby enhancing modelling and forecasting accuracy. This approach deepens our understanding of data patterns, improves predictions, and provides interpretability through component separation. However, some limitations persist when this type of forecasting model is used. Most time-frequency decomposition ensembles use a one-time decomposition approach, which leads to the inclusion of future information in the training data. Consequently, model inputs are influenced by in-sample and out-of-sample data, resulting in higher forecasting accuracy, as noted by Zhang et al. (2015) and Wu et al. (2022).

### 4.3. Discussion

In this subsection, we present a concise discussion of the diverse deep learning models used for price forecasting. Figure 19 illustrates the distribution of the deep learning models proposed in the reviewed studies. Notably, LSTM networks continue to hold a significant position in time series forecasting, both as key components of individual and ensemble models. This result suggests that capturing the temporal dependencies between data points is a priority for effective time series forecasting. In addition, the application of novel architectures, such as Transformers, GNNs, GANs, and DQNNs, for price forecasting is still in its early stages, mainly owing to the model complexity and high computational resources required for training. Moreover, neural architectures such as DNNs and 1D CNNs exhibit limited generalisation across different price forecasting tasks, despite their lower computational demands for training. Finally, we summarise the advantages and disadvantages of various deep learning models in Table 5.

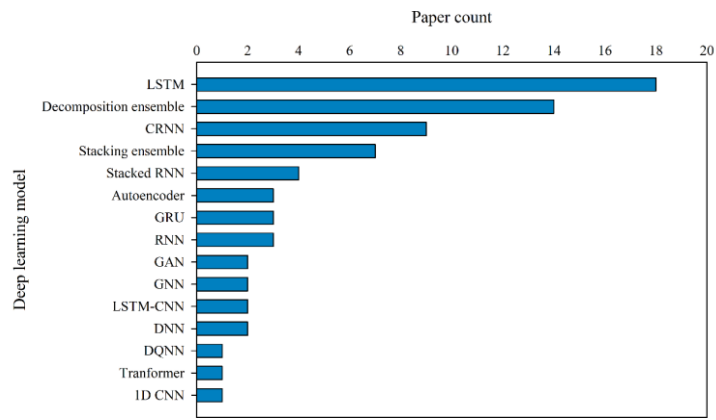

**Figure 19.** Distribution of deep learning models for price forecasting.

**Table 5.** Advantages and disadvantages of different deep learning models for price forecasting.

| DL model | Advantages | Disadvantages |
|---|---|---|
| **DNNs** | Can learn hierarchical representations, capturing complex abstractions. | Not explicitly designed for sequential data. Limited ability to capture temporal dependencies. |
| **1D CNNs** | Effective in capturing local dependencies and patterns in sequential data. Can learn hierarchical representations. | Struggle with modelling long-term dependencies. Limited applications for price forecasting. |
| **RNNs** | Widely applied and effective for capturing temporal dependencies in time series data. | Fixed memory capacity. May struggle with complex patterns or long-term dependencies. Susceptible to noise. |
| **CRNNs** | Capture both spatial features and temporal dependencies. Suitable for price forecasting tasks with spatiotemporal dynamics. | Computationally expensive. Hyperparameter tuning challenges. Interpretability may be difficult. |
| **Autoencoders** | Feature learning, capturing relevant patterns and relationships. | Lack of interpretability. Require sufficient training data. Risk of overfitting. |
| **Transformers** | Parallel processing, computational efficiency. Capturing complex patterns and dependencies. Incorporating diverse information. | Computational complexity for extremely long sequences. |
| **GNNs** | Capture complex dependencies and relationships between assets. Enhanced forecasting performance. | Construction of input graph can be challenging. Struggle with limited data or sudden market shifts. High computational requirements. |
| **GANs** | Capture complex patterns and dependencies. Generate realistic price sequences. | Require large amounts of data. Training instability. |
| **DQNNs** | Leverage the principles of quantum mechanics to formulate financial data dynamics. | Require specialised software and hardware. Challenging training and optimisation. |
| **Stacking Ensemble** | Combination of diverse models. Improved prediction accuracy. Adaptability to market conditions. | Additional computational resources. Increased model complexity. Interpretability challenges. |
| **Decomposition Ensemble** | Capture different components of time series. Improved understanding of patterns. Interpretable. | Exposure to future characteristics. Possible data leakage. |

## 5. FUTURE DIRECTIONS

Although diverse deep learning models have been applied to price forecasting, further research in this domain is still essential. Drawing from the insights gathered from the included studies, we propose several potential directions for future research in this area.

First, the application of deep learning models with complex structures to price forecasting requires further attention. The advancement of a large language model (LLM) based on the Transformer for generating and predicting new content implies that the Transformer is a potential successor of LSTM for time series forecasting. However, the application of Transformer models for price forecasting remains limited. The self-attention mechanism in a Transformer model can capture the temporal dependencies between distant positions in a sequence more effectively than LSTM models but requires quadratic time and memory space with respect to the input sequence length, making it less efficient for extremely long sequences. Therefore, additional research is required to examine the effectiveness of this type of deep learning model in various forecasting tasks. Similarly, the effectiveness of DQNNs for price forecasting is worth further investigation using more advanced software and hardware.

Second, extending deep learning models from point forecasting to interval forecasting in the financial sector is a promising direction for future research. In contrast to single-point forecasting, the prediction interval (PI) provides more robust and useful information because it not only estimates a point but also generates upper and lower bounds that encompass potential future values with a certain level of confidence. Several approaches to PI forecasting of time series have been proposed by Stankeviciute et al. (2021), Xu and Xie (2021), and Zaffran et al. (2022). These approaches can utilise point forecasting results generated from deep learning models to construct PIs of the target variable, which means that PI forecasting of financial time series may become a fruitful application area for various deep learning models.

Third, the reliability and validity of the predictions provided by the decomposition ensembles require further investigation. Most existing decomposition ensembles adopt one-time decomposition for the entire price series. Consequently, the decomposed components used to train the base models are derived from both in-sample and out-of-sample data, which may expose future characteristics and lead to higher forecasting accuracy (Zhang et al., 2015; Wu et al., 2022). Therefore, the effectiveness of these models in real-world forecasting requires further validation.

Fourth, the influence of data volume on model performance should be investigated in the context of price forecasting. For most of the studies reviewed herein, it was difficult to determine whether the quantity of data used for the model training was excessive or inadequate. It is imperative to emphasise that training with suitably sized datasets can potentially reduce the computational costs while obtaining optimal model performance. Research on this topic is highly valuable because time consumption and energy efficiency are important for the long-term application of forecasting models.

## 6. CONCLUSION

The increasing use of deep learning models for price forecasting in financial time series prompted us to conduct a literature review of related publications in the last three years. This review provides a comprehensive summary of deep learning models for price forecasting from two perspectives. From a theoretical perspective, we focused specifically on model architectures, which helps gain insights into how the model processes input data and makes decisions. From a practical perspective, we focused on summarising the applications of deep learning models in the reviewed studies to encourage further adoption of the proposed models. Although LSTM still plays a prominent role in building time series forecasting models, we observed that novel architectures, such as Transformers, GANs, GNNs, and DQNNs, have been adopted in price forecasting tasks, indicating that the financial sector truly benefits from recent advances in deep learning techniques. The insights and summaries provided in this review will aid researchers working in this rapidly developing interdisciplinary area.

Although our review covers the recent advancements in this field, it has several limitations. First, there is a possibility that relevant studies were not included in the selected databases. Although comprehensive keywords were used in the literature search, some studies may have been missed because they used less common linguistic terms. Second, although most of the included studies suggest that deep learning is a state-of-the-art technique for predicting financial time series, this review does not aim to provide an exhaustive experimental comparison between deep learning and other prediction methods, nor does it aim to make a comparison among the included studies. These comparisons would require significant computational resources and will be topics for future research.

Finally, potential directions for future research are presented. The first aspect that requires further attention is examining the effectiveness of deep learning models with complex structures for price forecasting. The second aspect pertains to extending the use of deep learning models from point forecasting to PI forecasting in the financial sector. In addition, the reliability and validity of the decomposition ensembles should be examined, and the influence of data quantity on model performance requires further exploration for long-term application of these models.

# Appendix

**Abbreviations**

| | | |
|---|---|---|
| ABC | - | Artificial Bee Colony Algorithm |
| ADARNN | - | Adaptive Recurrent Neural Network |
| AM | - | Attention Mechanism |
| ANN | - | Artificial Neural Network |
| AR | - | Auto Regressive |
| ARFIMA | - | Autoregressive Fractionally Integrated Moving Average |
| ARIMA | - | Autoregressive Integrated Moving Average |
| ARMA | - | Autoregressive Moving Average |
| BARDR | - | Bayesian ARD Regression |
| BERT | - | Bidirectional Encoder Representations from Transformer |
| BiGRU | - | Bidirectional Gated Recurrent Unit |
| BiLSTM | - | Bidirectional Long Short-Term Memory |
| BIRCH | - | Balanced Iterative Reducing and Clustering Using Hierarchy |
| BPNN | - | Back Propagation Neural Network |
| CEEMD | - | Complementary Ensemble Empirical Mode Decomposition |
| CEEMDAN | - | Complete Ensemble Empirical Mode Decomposition with Adaptive Noise |
| CNN | - | Convolutional Neural Network |
| ConvLSTM | - | Convolutional LSTM |
| CRNN | - | Convolutional-Recurrent Neural Network |
| CSI | - | Cubic Spline Interpolation |
| DA | - | Dual-Stage Attention-Based |
| DBGRUNN | - | Deep Bidirectional GRU Neural Network |
| dLSTM | - | Delayed Long Short-Term Memory |
| DP | - | Differential Privacy |
| DNN | - | Deep Neural Network |
| DQNN | - | Deep Quantum Neural Network |
| D-stat | - | Directional Statistical Indicator |
| DTR | - | Decision Tree Regressor |
| ECA | - | Efficient Channel Attention |
| EEMD | - | Ensemble Empirical Mode Decomposition |
| ELM | - | Extreme Learning Machine |
| EMD | - | Empirical Mode Decomposition |
| EMH | - | Efficient Market Hypothesis |
| ENet | - | Elastic Net |
| ENN | - | Elman Neural Network |
| ERNN | - | Elman Recurrent Neural Network |
| EWT | - | Empirical Wavelet Transform |
| FC | - | Fully Connected Layer |
| FFNN | - | Feed-forward Neural Network |
| FTD | - | Fourier Transform Denoising |
| GA | - | Genetic Algorithm |
| GAN | - | Generative Adversarial Network |
| GARCH | - | Generalised Autoregressive Conditional Heteroskedasticity |
| GCN | - | Graph Convolution Network |
| GNN | - | Graph Neural Network |
| GPR | - | Gaussian Process Regression |
| GPU | - | Graphics Processing Unit |
| GRNN | - | Generalised Regression Radial Basis Neural Network |
| GRU | - | Gated Recurrent Unit |
| GRUA | - | GRU Neural Network with Attention Mechanism |
| GRUNN | - | GRU Neural Network |
| IMF | - | Intrinsic Mode Function |
| KNN | - | K-Nearest Neighbors |
| KNR | - | K-Nearest Neighbor Regression |
| LLM | - | Large Language Model |
| LSTM | - | Long Short-Term Memory |
| LSTMRT | - | Long Short-Term Memory with Random Time Effective Function |
| LSSVR | - | Least Squares Support Vector Regression |
| MAE | - | Mean Absolute Error |
| MAAPE | - | Mean Arctangent Absolute Percentage Error |
| MAPE | - | Mean Absolute Percentage Error |
| MCCS | - | Multiscale Composite Complexity Synchronisation |
| MdAPE | - | Median of the Absolute Percentage Error |
| MedAE | - | Median Absolute Error |
| MEEMD | - | Modified Ensemble Empirical Mode Decomposition |
| MEMD | - | Multivariate Empirical Mode Decomposition |
| MLP | - | Multilayer Perceptron |
| MRC | - | Multi-scale Residual Convolutional Neural Network |
| MSE | - | Mean Square Error |
| MSLE | - | Mean Squared Log Error |
| NLP | - | Natural Language Processing |
| NRMSE | - | Normalisation Root Mean Squared Error |
| ORELM | - | Outlier Robust Extreme Learning Machine |
| PCA | - | Principal Component Analysis |
| PI | - | Prediction Interval |
| PR | - | Polynomial Regression |
| PSO | - | Particle Swarm Optimisation |
| PSR | - | Phase-Space Reconstruction |
| QNN | - | Quantum Neural Network |
| $R^2$ | - | Coefficient of Determination |
| RELM | - | Regular Extreme Learning Machine |
| RFR | - | Random Forest Regression |
| RIF | - | Random Inheritance Formula |
| RKELM | - | Robust Kernel Extreme Learning Machine |
| RMSE | - | Root Mean Square Error |
| RNN | - | Recurrent Neural Network |
| RSE | - | Root Relative Squared Error |
| SDAPE | - | Standard Deviation of the Absolute Percentage Error |
| SMAPE | - | Symmetric Mean Absolute Percentage Error |
| SSA | - | Singular Spectrum Analysis |
| SVM | - | Support Vector Machine |
| SVNN | - | Support Vector Neural Network |
| SVR | - | Support Vector Regression |
| SW | - | Stochastic Time Effective Weights |
| TIC | - | Theil Inequality Coefficient |
| VAE | - | Variational Autoencoder |
| ViT | - | Video Transformer |
| VMD | - | Variational Mode Decomposition |
| WNN | - | Wavelet Neural Network |
| WNNRT | - | Random Wavelet Neural Network |
| WPD | - | Wavelet Packet Decomposition |
| WT | - | Wavelet Transform |
| XGBoost | - | Extreme Gradient Boosting |

## Conflict of Interest

The authors declare that they have no conflicts of interest.

## Funding Information

This study did not receive any specific grants from funding agencies in the public, commercial, or non-profit sectors.

**Acknowledgments**

Not applicable.

**Notes**

Not applicable.